\documentclass[twocolumn,aps,prc,groupedaddress,longbibliography,showpacs,nofootinbib,floatfix]{revtex4-1}

\usepackage{amssymb,amsmath}
\usepackage{graphicx}
\usepackage{siunitx}
\usepackage{xspace}

\newcommand{\gev}[1]{\SI{#1}{\giga\electronvolt}}
\newcommand{\mb}[1]{\SI{#1}{\milli\barn}}

\newcommand{\sqs}{\sqrt{s}}
\newcommand{\snn}{\sqrt{s_{_{NN}}}}
\newcommand{\pt}{p_T}

\newcommand{\xf}{x_\textrm{F}}
\newcommand{\w}{W}
\newcommand{\wmin}{W_\textrm{min}}
\newcommand{\wmax}{W_\textrm{max}}

\newcommand{\egampr}{\omega^\textrm{rest}_{\gamma^\ast}}

\newcommand{\omegan}{\Omega_n}
\newcommand{\bmin}{b_\textrm{min}}

\newcommand{\an}{A_N}

\newcommand{\ep}{ep}

\newcommand{\gp}{\gamma^{\ast}p^{\uparrow}}
\newcommand{\gppin}{\gamma^{\ast}p^{\uparrow}\to\pi^+n}
\newcommand{\gppipin}{\gamma^{\ast}p^{\uparrow} \to \pi^{+}\pi^{0}n}

\newcommand{\unpp}{pp}
\newcommand{\pp}{p^{\uparrow}p}
\newcommand{\ppnX}{pp\to{nX}}
\newcommand{\polppnX}{p^{\uparrow}p\to{nX}}

\newcommand{\unpA}{pA}
\newcommand{\pA}{p^{\uparrow}A}
\newcommand{\pApin}{p^{\uparrow}A\to\pi^{+n}}
\newcommand{\pAnX}{pA\to{nX}}

\newcommand{\unpAu}{p\textrm{Au}}
\newcommand{\pAu}{p^{\uparrow}\textrm{Au}}
\newcommand{\pAupin}{p^{\uparrow}\textrm{Au}\to\pi^+n}
\newcommand{\pAupipin}{p^{\uparrow}\textrm{Au}\to\pi^{+}\pi^{0}n}
\newcommand{\pAunX}{p^{\uparrow}\textrm{Au}\to{nX}}

\newcommand{\unpAl}{p\textrm{Al}}
\newcommand{\pAl}{p^{\uparrow}\textrm{Al}}

\newcommand{\etal}{\textit{et al}.}

\newcommand{\onemaid}{\textsc{maid{\footnotesize{2007}}}}
\newcommand{\twomaid}{\textsc{\footnotesize{2}-pion maid}}

\begin{document}

\title{Recently measured large $\an$ for forward neutrons in $\pA$ collisions at
$\snn=200$\,GeV explained through simulations of ultraperipheral collisions and
hadronic interactions}

\author{Gaku Mitsuka}

\affiliation{RIKEN BNL Research Center, Brookhaven National Laboratory, 
Upton, New York 11973-5000, USA}

\date{\today}

%------------------------------------------------------------------------------|

% ---- Abstract ----
\begin{abstract}

The PHENIX experiment at the Relativistic Heavy Ion Collider recently reported
that transverse single spin asymmetry, $\an$, for forward neutrons in $\pA$
collisions at $\snn=\gev{200}$. $\an$ in $\pAl$ and $\pAu$ collisions were
measured as -0.015 and 0.18, respectively. These values
are clearly different from the measured $\an=-0.08$ in $\pp$ collisions.
In this paper, we propose that a large $\an$ for forward neutrons in
ultraperipheral $\pA$ collisions may explain the PHENIX measurements. The
proposed model is demonstrated using two Monte Carlo simulations. In the
ultraperipheral collision simulation, we use the \textsc{starlight} event
generator for the simulation of the virtual photon flux and then use the
\textsc{maid{\footnotesize{2007}}} unitary isobar model for the simulation of
the neutron production in the interactions of a virtual photon with a polarized
proton. In the $\pA$ hadronic interaction simulation, the differential cross
sections for forward neutron production are predicted by a simple one-pion
exchange model and the Glauber model. The simulated $\an$ values for both the
contribution of ultraperipheral collisions and the hadronic interactions are in
good agreement with the PHENIX results.

\end{abstract}

% ---- PACS number ----

\maketitle

%
% ----- Introduction -----
%
\section{Introduction}\label{sec:introduction}

The PHENIX experiment at the Relativistic Heavy Ion Collider (RHIC) reported
that the transverse single spin asymmetry, denoted $\an$, for forward neutrons
measured in transversely polarized proton--nucleus ($\pA$) collisions at
$\snn=\gev{200}$ is far different from that in polarized proton--proton ($\pp$)
collisions at $\sqs=\gev{200}$~\cite{PHENIXPrelim}. That is, the measured $\an$
for $\pAl$ and $\pAu$ collisions are $-0.015 \pm 0.005$ and $0.18 \pm 0.03$,
respectively. On the other hand, $\an$ is $-0.08\pm0.02$ in $\pp$
collisions~\cite{PHENIXNeutron}. In these measurements, the neutrons produced in
$\pp$ and $\pA$ collisions are detected by a zero-degree
calorimeter~\cite{RHICZDC} in the polarized proton remnant side that is defined
as the positive rapidity region.

As studied in Ref.~\cite{Kopeliovich2}, the interference of pion (spin-flip) and
$a_1$-Reggeon (spin nonflip) exchanges successfully explains $\an$ for forward
neutrons in $\pp$ collisions at RHIC. We would expect that this mechanism can
be extended to predict $\an$ for $\pA$ collisions. The authors of
Ref.~\cite{Kopeliovich2} incorporate the pion--$a_1$-Reggeon interference with
strong nuclear absorptive corrections and a nuclear breakup~\cite{Kopeliovich3}.
However, the predicted $\an$ in $\pA$ collisions remains negative and the
magnitude of $\an$ is too small to explain the PHENIX results.

In this paper, we propose an alternative mechanism: ultraperipheral
proton--nucleus collisions (UPCs)~\cite{Bertulani1,Bertulani2}. UPCs occur when
the impact parameter $b$ is larger than the sum of the radii of each colliding
particle, namely, $b > R_p + R_A$, where $R_p$ and $R_A$ are the radius of the
proton and nucleus, respectively.
As we explored in Ref.~\cite{Mitsuka}, UPCs have a comparable cross section with
the hadronic interactions in the very forward rapidity region. In
ultraperipheral $\pA$ collisions, virtual photons ($\gamma^\ast$) emitted from
the relativistic nucleus interact with polarized protons.
Thus the number of neutrons produced via the $\gp$ interaction depends on the
azimuthal angle of the scattered neutrons relative to the proton polarization.
This may finally contribute to the large $\an$ for forward neutrons in $\pA$
collisions.

We first develop the Monte Carlo (MC) simulation framework for UPCs
(Sec.~\ref{sec:upcsim}) and hadronic interactions (Sec.~\ref{sec:qcdsim}). In
Sec.~\ref{sec:results}, using these MC simulations we show that UPCs in $\pA$
collisions have sizable cross sections compared with hadronic interactions and
the yield of forward neutrons in UPCs certainly depend on the scattering
azimuthal angle relative to the proton polarization axis. Finally, in
Sec.~\ref{sec:discussions}, we compare the simulated $\an$ with the PHENIX
results.  Conclusions are drawn in Sec.~\ref{sec:conclusions}.
Natural units $\hbar=c=1$ are used throughout.

%
%---- UPC MC simulation ----
%
\section{Methodology of Ultraperipheral Collisions Monte Carlo simulations}
\label{sec:upcsim}

The MC simulation for UPCs in this study comprises two steps. First, we simulate
the virtual photon flux as a function of the photon energy and impact parameter
by using \textsc{starlight}~\cite{STARLIGHT,STARLIGHTcode}
(Sec.~\ref{sec:photon}). Second, the simulation of the $\gp$ interaction and
particle production is performed following the differential cross sections that
are predicted by the $\onemaid$ model~\cite{MAID07} (Sec.~\ref{sec:photon+p}).

\subsection{Formalism for ultraperipheral $\pA$ collisions}
\label{sec:UPC}

The differential cross section for single neutron production in $\pA$ UPCs is
given by
\begin{equation}
\frac{d\sigma_{\textrm{UPC}(\pApin)}^4}{d\w db^2 d\omegan} = 
\frac{d^3N_{\gamma^\ast}}{d\w db^2}
\frac{d\sigma_{\gppin}(\w)}{d\omegan}
\overline{P_\textrm{had}}(b)
\label{eq:dsigupc}
\end{equation}
where $N_{\gamma^\ast}$ is the number of the emitted photons, $\w$ is the $\gp$
center-of-mass energy, $d\omegan=\sin\theta_n\,d\theta_n\,d\phi_n$ with the
neutron scattering polar angle $\theta_n$ and azimuthal angle $\phi_n$ in the
$\gp$ center-of-mass frame, $\sigma_{\gppin}(\w)$ is the total cross section for
a single photon interaction with a proton leading to single neutron production,
and $\overline{P_\textrm{had}}(b)$ is the probability of having no hadronic
interactions in $\pA$ collisions at given $b$. We calculate
$d^3N_{\gamma^\ast}/d\w db^2$ and $d\sigma_{\gppin}(\w)/d\omegan$ in
Sec.~\ref{sec:photon} and Sec.~\ref{sec:photon+p}, respectively.

A finite probability for having no hadronic interactions
$\overline{P_\textrm{had}}(b)$ is introduced in order to account for a smooth
cut off around the impact parameter $b=R_p + R_A$~\cite{STARLIGHT}. $R_p$ and
$R_A$, where $R_\textrm{Al}\sim\SI{5}{\femto\meter}$ and
$R_\textrm{Au}\sim\SI{7}{\femto\meter}$, are the radius of the proton and
nucleus, respectively. The range of the impact parameter considered in the
simulation extends from $b_\textrm{\rm min} = \SI{4}{\femto\meter}$ to
$b_\textrm{\rm max} = \SI{e5}{\femto\meter}$.
The value of $b_\textrm{\rm min}$ is well below the sum of the effective radii
of colliding particles, and $\overline{P_\textrm{had}}(b)$ rapidly approaches
zero below these nuclear radii.

\begin{figure*}[!tbh]
  \includegraphics[width=0.48\linewidth]{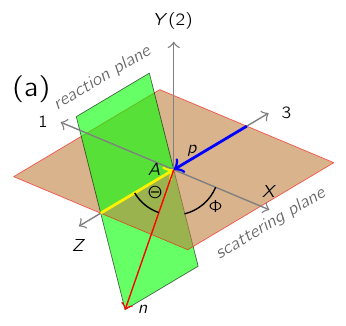}
  \includegraphics[width=0.48\linewidth]{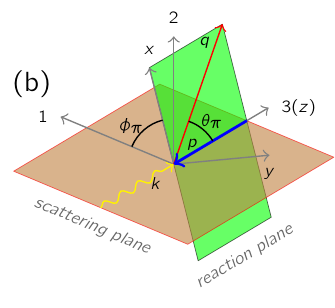}  
\caption{(a) coordinate axes in the detector reference frame. (b) kinematical
variables and coordinate axes in the $\gp$ center-of-mass frame $\{0,2,3\}$.}
  \label{fig:frame}
\end{figure*}

For comparisons with the simulation results of hadronic interactions in
Sec.~\ref{sec:results} and the experimental results from PHENIX in
Sec.~\ref{sec:discussions}, we will numerically transform the differential cross
section in Eq.~(\ref{eq:dsigupc}) based on the $\gp$ center-of-mass frame to
that in the detector reference frame. Both frames are defined in
Fig.~\ref{fig:frame}.

%----------------------------------------------------- Fig_1

\subsection{Simulation of the virtual photon flux}
\label{sec:photon}

For simplicity (unless otherwise noted), the discussion in this subsection is
based on the proton rest frame. The virtual photons flux emitted by the
relativistic nucleus is simulated using \textsc{starlight}, which follows the
Weizs\"{a}cker-Williams approximation~\cite{Weizsacker, Williams}. The double
differential photon flux due to the fast moving nucleus with velocity $\beta$ is
written as
\begin{equation} \frac{d^3N_{\gamma^\ast}}{d\egampr db^2} =
\frac{Z^2\alpha}{\pi^2}\frac{x^2}{\egampr b^2} \left( K_1^2(x) +
\frac{1}{\gamma^2}K_0^2(x) \right), \label{eq:gflux}
\end{equation} 
where $\egampr$ is the photon energy, $Z$ is the atomic number ($Z = 13$ and
$79$ for Al and Au, respectively), $\alpha$ is the fine structure constant, $x =
\egampr b/\gamma$ ($\gamma = \sqrt{1-\beta^2}^{-1/2}$ is the Lorentz factor),
and $K_0$ and $K_1$ are the modified Bessel functions. In the case of a
relativistic nucleus ($\gamma \gg 1$), we safely disregard the contribution of
the term $K_0^2(x)/\gamma^2$ in Eq.~(\ref{eq:gflux}). The photon energy
$\egampr$ in the proton rest frame is properly transformed to the $\gp$
center-of-mass energy $\w$ to allow substitution of Eq.~(\ref{eq:gflux}) for
Eq.~(\ref{eq:dsigupc}).

\subsection{Simulation of the $\gp$ interaction}
\label{sec:photon+p}

The kinematics of the $\gp$ interaction is shown in Fig.~\ref{fig:frame}(b) and
is defined as
\begin{equation}
\gamma^{\ast}\,(k) + p^{\uparrow}\,(p) \to \pi^+\,(q) + n\,(n),
\label{eq:kinematics}
\end{equation}
where the variables in brackets indicate the four-momenta of each particle. We
use the following notations for these four-momenta:
\begin{eqnarray}
k^\mu &=& (\omega_{\gamma^\ast},\boldmath{k}),  \\ \nonumber
p^\mu &=& (\omega_p,  -\boldmath{k}),           \\ \nonumber
q^\mu &=& (\omega_\pi, \boldmath{q}),           \\ \nonumber
n^\mu &=& (\omega_n,  -\boldmath{q}),
\label{eq:notation}
\end{eqnarray}
where the $\gp$ center-of-mass energy is given by
$\w=\omega_{\gamma^\ast}+\omega_{p^{\uparrow}}$.

Figure~\ref{fig:frame}(b) also introduces the polar angle $\theta_\pi$ and
azimuthal angle $\phi_\pi$ of $\bm{q}$, with reference to a coordinate system
with $\bm{k}$, 1, and 2 axes such that $\bm{k}$ lies in the 1-3 plane. The
proton is transversely polarized along the 2 axis.  The frame $\{1,2,3\}$ is the
scattering plane. The frame $\{x,y,z\}$ is defined such that the $z$ axis is
directed into the $\bm{k}$ direction; the $y$ axis is perpendicular to the
$x$--$z$ reaction plane; and the $x$ axis is given by $x=y{\times}z$.

Single neutron and pion production from the $\gp$ interaction are simulated
following the differential cross sections predicted by the $\onemaid$ model. The
cross section of the $\gppin$ interaction is formed as in Ref.~\cite{Drechsel}:
\begin{eqnarray}
\frac{d\sigma_{\gppin}}{d\Omega_\pi} &=&
\frac{|\bm{q}|}{\omega_{\gamma^\ast}}(R_T^{00} + P_y R_T^{0y}) \nonumber \\
&=& \frac{|\bm{q}|}{\omega_{\gamma^\ast}}R_T^{00}
(1 + P_2 \cos\phi_\pi T(\theta_\pi)),
\label{eq:gpxsec}
\end{eqnarray}
where $R_T^{00}$ and $R_T^{0y}$ are the response functions for pion
photoproduction, and $P_y$ and $P_2$ are the proton polarization along $y$ and
$2$ axes, respectively. In the third equation, $R_T^{0y}/R_T^{00}$ and $P_y$
respectively are replaced with target asymmetry $T(\theta_\pi)$ and $P_2
\cos\phi_\pi$. We assume $P_2 = 1$ and $P_1 = 0$ in this study. Note that we can
numerically obtain $d\sigma_{\gppin}/d\Omega_n$ in Eq.~(\ref{eq:dsigupc}) from
$d\sigma_{\gppin}/d\Omega_\pi$ in Eq.~(\ref{eq:gpxsec}) with the relation
$\theta_n=\pi-\theta_\pi$ and $\phi_n=\pi-\phi_\pi$.

The photon virtuality is limited by $Q^2<(1/R_A)^2$, thus
$Q^2<\SI{2e-3}{\giga\electronvolt^{2}}$ in $\pAl$ collisions and
$Q^2<\SI{6e-4}{\giga\electronvolt^{2}}$ in $\pAu$ collisions. In the following
simulations, we fix $Q^2=\SI{0}{\giga\electronvolt^{2}}$ as default and take
into account effects of nonzero $Q^2$ as a model uncertainty. The virtual
photons are assumed to be fully unpolarized in this study.

Both baryon resonance and nonresonance contributions are taken into account for
neutron and pion productions in $\onemaid$. The model contains all four-star
resonances with masses below \gev{2} and follows the Breit-Wigner forms for the
resonance shape. Nonresonance background contribution contains the Born terms as
detailed in Ref.~\cite{MAID07}.

Due to the energy range guaranteed by $\onemaid$, the minimum and maximum $\w$
values in the UPC MC simulation are set as $\wmin=\gev{1.1}$ and
$\wmax=\gev{2.0}$, respectively. We will discuss cross section contribution from
outside the above energy range in Sec.~\ref{sec:$p$$+$Au}.

\begin{figure}[!htb]
  \includegraphics[width=1.0\linewidth]{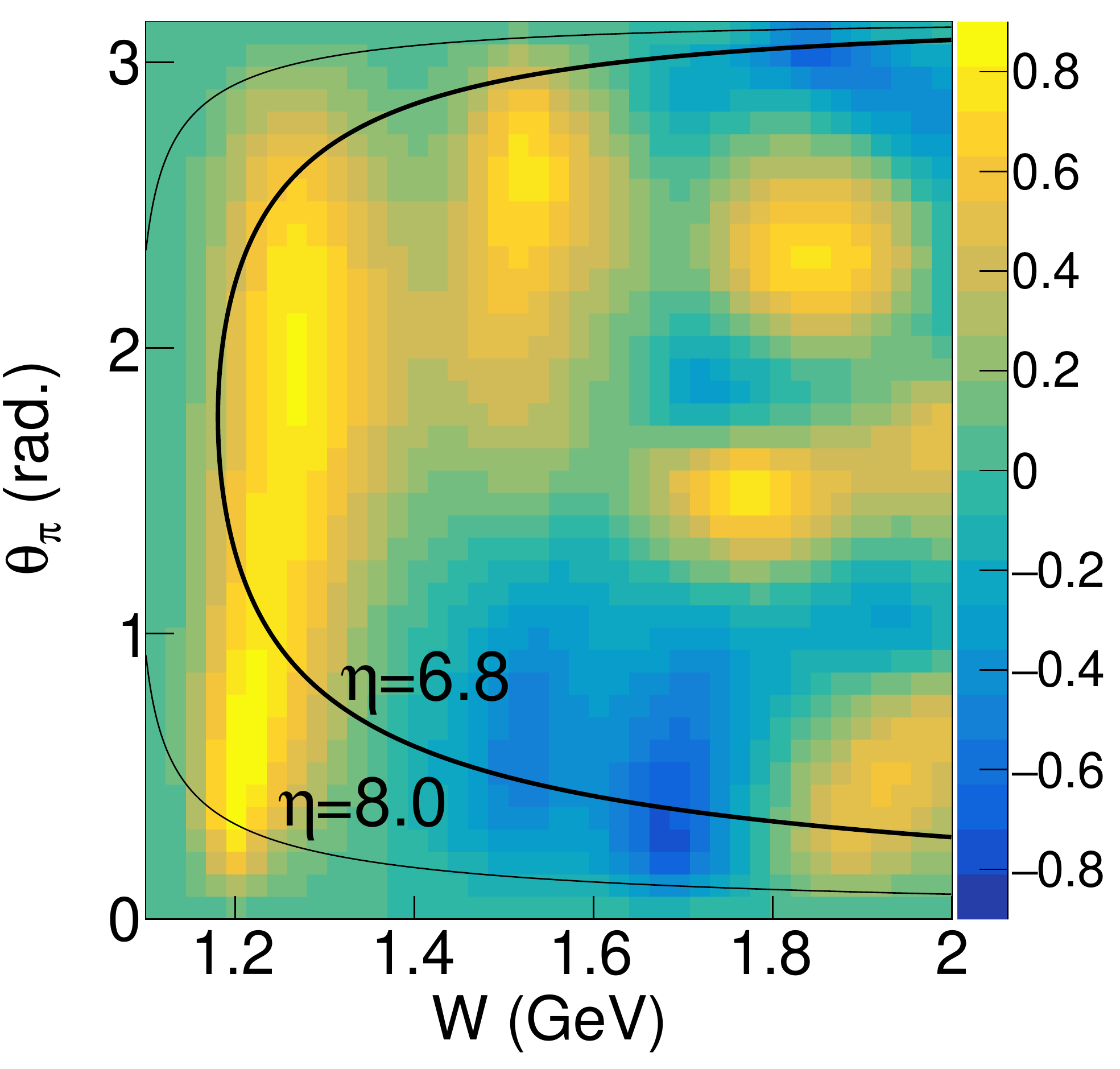}
\caption{Target asymmetry $T(\theta_\pi)$ of the $\gppin$ interaction as
function of $\w$. In the detector reference frame, the thick and thin curves
correspond to the rapidity of produced neutrons $\eta=6.8$ and 8.0,
respectively.}
  \label{fig:T}
\end{figure}

Figure~\ref{fig:T} shows $T(\theta_\pi)$ of the $\gppin$ interaction as a
function of $\w$. In the detector reference frame, the thick and thin curves
correspond to the rapidity of produced neutrons of $\eta=6.8$ and 8.0,
respectively.
Because the neutrons produced along the negative $z$ axis, namely, $\theta_\pi
\sim 0$, are more likely to have large positive rapidity in the detector
reference frame, the thick and thin curves shift to the large $\theta_\pi$
region.

%-------------------------------------------------- Fig_2

%
%---- QCD MC simulation ----
%
\section{Methodology of simulations in Hadronic interactions}
\label{sec:qcdsim}

Throughout this section, we use the detector reference frame defined in
Fig.~\ref{fig:frame}(a).  We effectively obtain the differential cross section
of forward neutron production in $\pA$ hadronic interactions as follows.  First
(Sec.~\ref{sec:$p$$+$$p$}), we calculate the cross section of inclusive neutrons
in $\unpp$ collisions, $\sigma_{\ppnX}$, using a simple one-pion exchange model.
Note that this calculation is performed for unpolarized protons. The one-pion
exchange model has well described forward neutron production in $\unpp$
collisions at the Intersecting Storage Rings~\cite{ISR} and
RHIC~\cite{PHENIXNeutron} and in $\ep$ collisions at the Hadron-Electron Ring
Accelerator (HERA)~\cite{ZEUS1}.

Second (Sec.~\ref{sec:glauber}), the cross section of the $\pAnX$ interaction,
$\sigma_{\pAnX}$, can be calculated by $\sigma_{\ppnX}$ and the Gribov-Glauber
model~\cite{Gribov,Glauber}. Here we avoid an implementation of multiple
scattering of a projectile proton with a nucleus, and instead we multiply the
$\unpp$ cross section $\sigma_{\ppnX}$ with the inelastic cross section ratio
$\sigma_{\unpA}/\sigma_{\unpp}$ obtained from Ref.~\cite{Guzey}.

For the $\Phi$-dependence of the differential cross section, we multiply
$d\sigma_{\pAnX}/d\omegan$ with $1+\cos\Phi \an^{\textrm{HAD}(\unpA)}$ in order
to effectively take into account the single spin asymmetry
$\an^{\textrm{HAD}(\unpA)}$ (Sec.~\ref{sec:glauber}).

\subsection{Simulation of the $\unpp$ interaction}
\label{sec:$p$$+$$p$}

The differential cross section for inclusive neutrons in $\unpp$ collisions at
the center-of-mass energy $\sqs$ as a function of the longitudianl momentum
fraction $\xf$ and the transverse momentum $\pt$ is formed in terms of the
pion-exchange model~\cite{Kopeliovich1} as
\begin{eqnarray}
\xf\frac{d\sigma_{\ppnX}}{d\xf d\pt^2} &=&
S^2\left(\frac{\alpha'_\pi}{8}\right)^2 |t|G^2_{\pi^+pn}(t)|\eta_\pi(t)|^2 \nonumber \\
&\times& (1-\xf)^{1-2\alpha_\pi(t)}\sigma^\textrm{\rm tot}_{\pi^++p}(M_X^2),
\label{eq:dsighadpp}
\end{eqnarray}
where $S^2$ is the rapidity gap survival factor,
$\alpha_\pi=\alpha'_\pi(t-m_\pi^2)$ is the pion trajectory with the slope
$\alpha'_\pi$ and the pion mass $m_\pi$, $t$ is the four-momentum transfer
squared, $G_{\pi^+pn}(t)$ is the effective vertex function, $\eta_\pi(t)$ is the
phase factor~\cite{Kopeliovich1}, and $\sigma^\textrm{\rm tot}_{\pi^+p}(M_X^2)$
is the total cross section of the $\pi^+p\to{X}$ interactions at the $\pi^+p$
center-of-mass energy $M_X^2=(1-\xf)s$. The effective vertex function is
parameterized as $G_{\pi^+pn}(t)=g_{\pi^+pn} e^{R^2_\pi t}$ using the
pion--nucleon coupling $g_{\pi^+pn}$ and the $t$-slope parameter $R^2_\pi$. In
this study, we fix $\alpha'_\pi=\SI{1.0}{\per\square\giga\electronvolt}$ and
$g^2_{\pi^+pn}/8\pi=13.75$ which are consistent with the results at
HERA~\cite{ZEUS1,H1}, and follow the best {\sc compete} fit results~\cite{PDG}
for $\sigma^\textrm{\rm tot}_{\pi^+p}(M_X^2)$. Because the parameters $S^2$ and
$R^2_\pi$ have been poorly determined to date, we use $S^2=0.2$ and
$R^2_\pi=\SI{0.3}{\per\square\giga\electronvolt}$ that derive the best agreement
with the forward neutron $d\sigma_{\polppnX}/d\xf$ distribution measured at the
PHENIX experiment~\cite{PHENIXNeutron}. These best-fit values are compatible
with other experimental results~\cite{ZEUS2}.

\subsection{Single spin asymmetry in $\pA$ collisions}
\label{sec:glauber}

As introduced in the third paragraph of Sec.~\ref{sec:qcdsim}, we avoid an
implementation of multiple scattering of a projectile proton with a nucleus. On
the other hand, we effectively obtain the $\unpA$ cross sections
$\sigma_{\pAnX}$ by multiplying $\sigma_{\ppnX}$ in Eq.~(\ref{eq:dsighadpp})
with the inelastic cross section ratio $\sigma_{\unpA}/\sigma_{\unpp}=A^{0.42}$
that is calculated in Ref.~\cite{Guzey}. Thus we obtain:
\begin{equation}
\xf\frac{d\sigma_{\unpA \to nX}}{d\xf d\pt^2} = 
\xf\frac{d\sigma_{\unpp \to nX}}{d\xf d\pt^2} A^{0.42}.
\label{eq:dsighadpA}
\end{equation}

The single spin asymmetry for forward neutrons in $\pp$ interactions 
originate in the interference of pion 
(spin-flip) and $a_1$-Reggeon (spin nonflip) exchanges~\cite{Kopeliovich2} 
that well reproduces the result from the PHENIX experiment: 
$\an^{\unpp}=-0.08 \pm 0.02$~\cite{PHENIXNeutron}. Preliminary results 
in Ref.~\cite{Kopeliovich3} based on the same approach as 
Ref.~\cite{Kopeliovich2} state that single spin asymmetry for forward 
neutrons in hadronic $\pA$ collisions is also described by the 
pion--$a_1$-Reggeon interference followed by a nuclear breakup. 
Here, we do not implement the pion--$a_1$ interference in the 
simulation.  Instead, we multiply the $\unpA$ differential cross 
section in Eq.~(\ref{eq:dsighadpA}) by 
$1+\cos\Phi\an^{\textrm{HAD}(\unpA)}$, where we take 
$\an^{\textrm{HAD}(\unpAu)}=-0.05$ and 
$\an^{\textrm{HAD}(\unpAl)}=-0.05$ from Ref.~\cite{Kopeliovich3}.

Finally, we obtain using Eq.~(\ref{eq:dsighadpA}):
\begin{eqnarray}
\xf\frac{d\sigma_{\pA \to nX}}{d\xf d\pt^2}
&=& \xf\frac{d\sigma_{\unpA \to nX}}{d\xf d\pt^2}(1+\cos\Phi\an^{\textrm{HAD}(\unpA)}) \nonumber \\
&=& \xf\frac{d\sigma_{\unpp \to nX}}{d\xf d\pt^2} A^{0.42} \nonumber \\
&&\times (1+\cos\Phi\an^{\textrm{HAD}(\unpA)}).
\label{eq:dsighadpolpA}
\end{eqnarray}

%
%---- Results ----
%
\section{Results}
\label{sec:results}

\subsection{Simulation results in $\pAu$ collisions at $\snn=\gev{200}$}
\label{sec:$p$$+$Au}

%--- cross section
\subsubsection{The total cross sections}\label{sec:$p$$+$Au_xsec}

First, we calculate the total cross section of the $\pAu \to nX$ 
interaction at $\snn=\gev{200}$ and compare it with UPCs and hadronic 
interactions. The total cross section for UPCs is calculated by 
integrating Eq.~(\ref{eq:dsigupc}) over $W$, $b$, and $\omegan$:
\begin{eqnarray}
\sigma_{\textrm{UPC}(\pAupin)} &=&
\int_{\omegan}
\int_{\bmin}^{\bmin}
\int_{\wmin}^{\wmax}
\frac{d\sigma_{\textrm{UPC}(\pAupin)}^4}{d\w db^2 d\omegan} \nonumber \\
&\times& \overline{P_\textrm{had}}(b)2\pi b\,dW\,db\,d\omegan,
\label{eq:sigupc}
\end{eqnarray}
where we require a single neutron scattered at $y>6.9$ and $\xf>0.4$. The
rapidity limit corresponds to the acceptance of a zero-degree calorimeter at
RHIC and the $\xf$ limit is introduced to remove the contribution of low-energy
forward neutrons. These cuts are consistent with the RHIC
measurements~\cite{PHENIXNeutron}. As addressed in Sec.~\ref{sec:UPC}, we fix
$b_\textrm{\rm min} = \SI{4}{\femto\meter}$, $b_\textrm{\rm max} =
\SI{e5}{\femto\meter}$, $\wmin=\gev{1.1}$, and $\wmax=\gev{2.0}$. We then obtain
$\sigma_{\textrm{UPC}(\pAupin)}=\mb{19.6}$.

\begin{figure}[!htb]
  \includegraphics[width=1.0\linewidth]{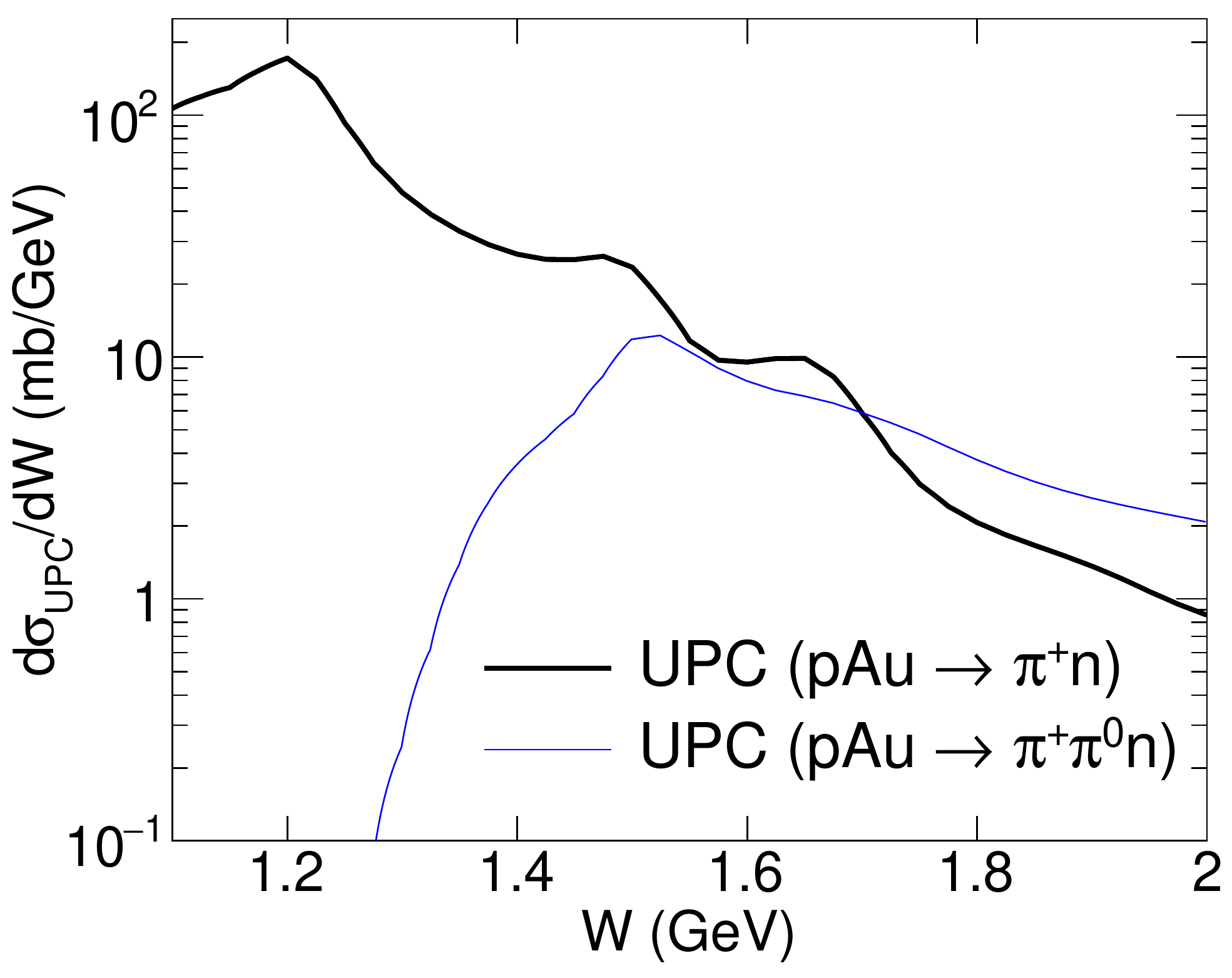}
\caption{The differential UPC cross sections as a function of $\w$.
Thick black curve indicates the $\pAupin$ interaction and thin blue curve
indicates the $\pAupipin$ interaction.}
  \label{fig:dsdw}
\end{figure}

For the discussions in Sec.~\ref{sec:$p$$+$Au_z} and \ref{sec:$p$$+$Au_sys},
here we show the differential UPC cross sections at $\snn=\gev{200}$ as a
function of $\w$ in Fig.~\ref{fig:dsdw}. The $d\sigma_{\textrm{UPC}(\pAu)}/d\w$
values are calculated by integrating Eq.~(\ref{eq:dsigupc}) over $b$ and
$\omegan$. For simplicity, no kinematical limit is applied to such integration.
Thick black curve indicates the $\pAupin$ interaction and thin blue curve
indicates the two-pion production $\gppipin$.

The total cross section for hadronic interaction is calculated by integrating
Eq.~(\ref{eq:dsighadpolpA}) over $\xf$ and $\pt$:
\begin{equation}
\sigma_{\textrm{HAD}(\pAunX)} = 2\pi\int_{\xf} \int_{\pt}
\frac{d\sigma_{\pAunX}}{d\xf d\pt^2}\pt\,d\pt\,d\xf.
\label{eq:sighad}
\end{equation}
We obtain $\sigma_{\textrm{HAD}(\pAunX)}=\mb{19.2}$ by requiring a single
neutron emitted into $y>6.9$ and $\xf>0.4$. According to the comparison of these
two cross sections, we find that UPCs lead to significant background
contribution to the investigations of single spin asymmetry in terms of hadronic
interactions. Table~\ref{tbl:singlexsec} summarizes the calculated cross
sections.

%=================================================== Table_I
\begin{table}[tbp]
\caption{Cross sections for neutron production in ultra-peripheral collisions
and hadronic interactions at $\snn=\gev{200}$. Cross sections in parentheses are
calculated without $\eta$ and $\xf$ limits.}
    \begin{ruledtabular}
	\begin{tabular}{cccc}
     \multicolumn{2}{c}{UPCs} &
     \multicolumn{2}{c}{Hadronic interactions} \\
     \cline{1-2}     \cline{3-4}
     $\pAl$                   & $\pAu$                     & $\pAl$     & $\pAu$      \\
     $\mb{0.7}$\,($\mb{2.2}$) & $\mb{19.6}$\,($\mb{41.7}$) & $\mb{8.3}$ & $\mb{19.2}$ \\
	\end{tabular}
	\end{ruledtabular}
\label{tbl:singlexsec}
\end{table}

%-------------------------------------------------- Fig_3
 
%--- ds/dz

\subsubsection{The differential cross sections as a function of 
$\xf$}\label{sec:$p$$+$Au_z}

In Fig.~\ref{fig:dsdxf}(a), we show the differential cross sections as a
function of $\xf$, namely, $d\sigma/d\xf$, for UPCs (dashed [red] line) and
hadronic interactions (solid [black] line). UPCs dominate in $d\sigma/d\xf$ at
$\xf>0.6$ and have a sharp peak around $\xf=0.95$.  This peak originates from
the $\gp \to \Delta^+(1232) \to \pi^{+}n$ channel in UPCs. As found in the thick
black curve in Fig.~\ref{fig:dsdw}, a $\gp$ center-of-mass energy of $1.1 < \w <
\gev{1.3}$, a photon energy ranging from $0.17 < \egampr < \gev{0.5}$ in the
proton rest frame, corresponds to the $\Delta^+(1232)$ baryon-resonance region
that has a larger UPC cross section compared to higher energy regions due to the
both ample photon flux and large $\gp\to\Delta^+(1232)$ cross section. Thus, low
momentum neutrons produced by a pronounced $\gp$ interaction at $1.1 < \w <
\gev{1.3}$ and emitted into $\theta_n\sim\pi$ in the $\gp$ center-of-mass frame
are boosted to nearly the same velocity of the projectile proton in the detector
reference frame. These neutrons lead to the forward neutrons sharply distributed
around $\xf=0.95$. Similarly, the neutrons emitted into $\theta_n\sim0$ at $1.1
< \w < \gev{1.3}$ in the $\gp$ center-of-mass frame cause the second peak round
$\xf=0.65$.

%--- ds/dPhi
\subsubsection{The differential cross sections as a function of $\Phi$}
\label{sec:$p$$+$Au_Phi}

\begin{figure*}[!htb]
  \includegraphics[width=0.998\linewidth]{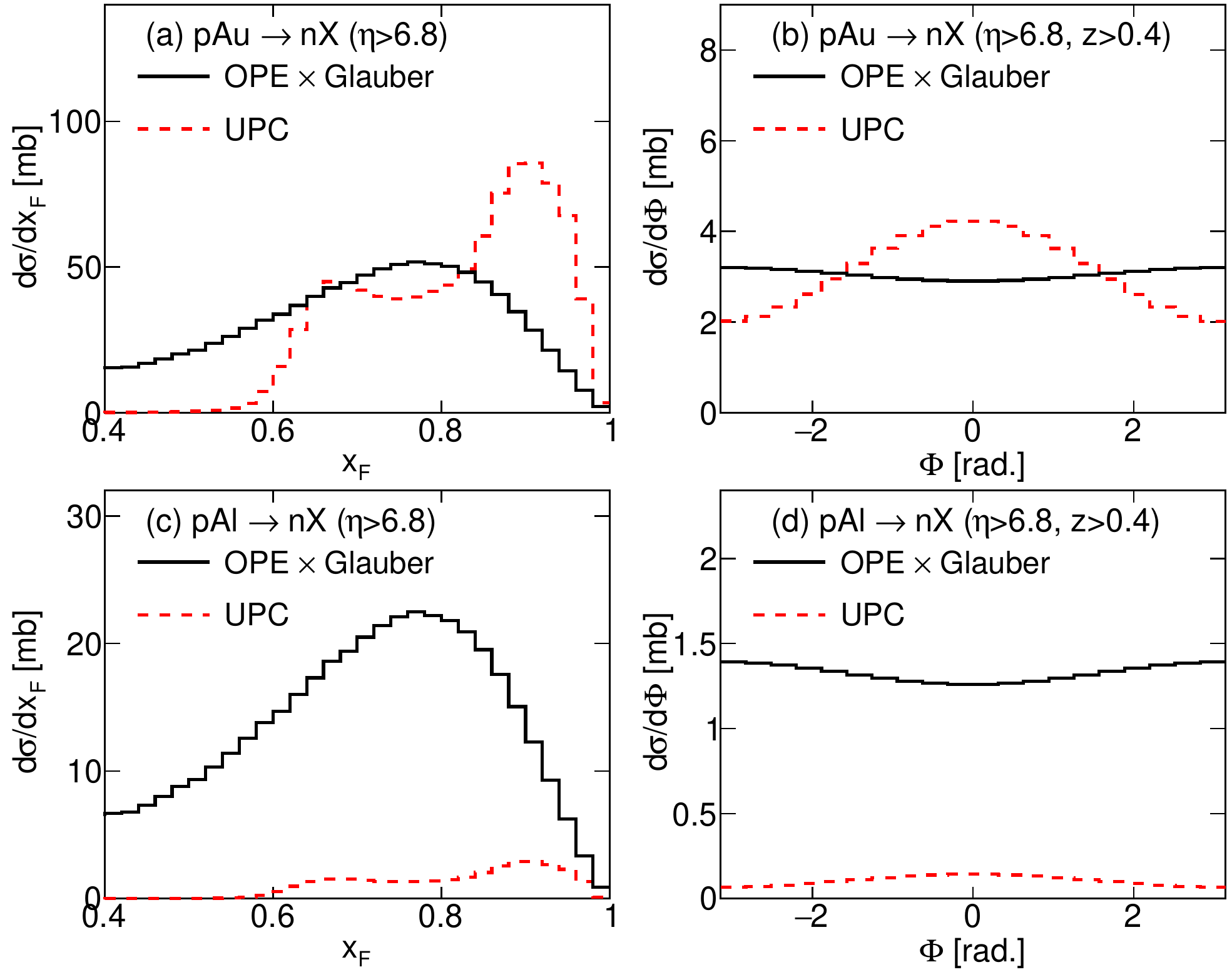}
\caption{The differential cross sections of UPCs and hadronic 
interactions as a function of $\xf$ and $\Phi$. Solid [black] lines 
indicate hadronic interactions and dashed [red] lines indicate UPCs.}
  \label{fig:dsdxf}
\end{figure*}

In Fig.~\ref{fig:dsdxf}(b), we compare the differential cross section as a
function of $\Phi$, namely, $d\sigma/d\Phi$ between UPCs (dashed [red] line) and
hadronic interactions (solid [black] line). We find that $d\sigma/d\Phi$ of UPCs
has substantial positive asymmetry $\an^{\textrm{UPC}(\unpAu)}$ of about 0.36
compared with the negative asymmetry of hadronic interactions
$\an^{\textrm{HAD}(\unpAu)}=-0.05$.

The UPC-induced asymmetry can be understood as follows. Replacing $\phi_\pi$
with $\Phi$ and $P_2=1$ in Eq.~(\ref{eq:gpxsec}), the $\Phi$-dependence of the
differential UPC cross section is approximated as
\begin{equation}
\frac{d\sigma_\textrm{UPC}}{d\Phi}
\propto 1+\cos\Phi \langle T(\theta_\pi) \rangle,
\label{eq:dsdPhi}
\end{equation}
where $\langle T(\theta_\pi) \rangle$ is an average of $T(\theta_\pi)$ over
$\theta_\pi$ but the rapidity and $\xf$ limits, $\eta<6.8$ and $\xf>0.4$, are
applied. As we find in the $d\sigma_{\textrm{UPC}(\pAu)}/d\w$ distribution in
Fig.~\ref{fig:dsdw}, forward neutrons in UPCs are mainly produced by the
$\Delta^+(1232) \to \pi^+ n$ decay at $1.1<\w<\gev{1.3}$, where $\langle
T(\theta_\pi) \rangle$ is $\sim0.7$, as shown in Fig.~\ref{fig:T}. Conversely,
resonances at $1.4<\w<\gev{1.8}$ have negative $\langle T(\theta_\pi) \rangle$
below $\theta_\pi\sim0.5$. Therefore $d\sigma_\textrm{UPC}/d\Phi$ integrating
over $\w$ suffers from the both positive and negative $\langle T(\theta_\pi)
\rangle$ and then we obtain $\langle T(\theta_\pi) \rangle=0.36$ at
$1.1<\w<\gev{2.0}$. In accordance with an equivalence $\an^{\textrm{UPC}(\unpA)}
= \langle T(\theta_\pi) \rangle$, we finally obtain
$\an^{\textrm{UPC}(\unpAu)}=0.36$.

%--- systematic errors
\subsubsection{Model uncertainties}
\label{sec:$p$$+$Au_sys}

Finally, we discuss the following three uncertainties in the present UPC 
cross sections: (1) the contribution from outside $1.1<\w<\gev{2.0}$, 
(2) the contribution from the two-pion production process, and 
(3) the effects of nonzero $Q^2$.

(1) We first compare the UPC cross sections in the following three 
energy ranges: $\w<\gev{1.1}$, $1.1<\w<\gev{2.0}$, and $\w>\gev{2.0}$. 
For the calculation of UPC cross sections, we use the framework in 
Ref.~\cite{Mitsuka} instead of the framework developed in this paper, 
because $\onemaid$ provides the $\gp$ differential cross sections only at 
$1.1<\w<\gev{2.0}$. In the framework in Ref.~\cite{Mitsuka}, the proton 
polarization is not taken into account, however the cross sections 
integrated over polar and azimuthal angles are independent of the target 
polarization. Unlike the framework developed in this paper, the total 
$\gp$ cross section $\sigma_{\gppin}(\w)$ in Ref.~\cite{Mitsuka} is 
taken from the compilation of present experimental results~\cite{PDG} at 
$\w<\gev{7}$ and from the best {\sc compete} fit results~\cite{PDG} at 
$\w>\gev{7}$. The UPC cross sections in each energy range are summarized 
in Table.~\ref{tbl:singlexsecsys}. Note that the rapidity and $\xf$ 
limits, $\eta>6.9$ and $\xf>0.4$, are applied to the these cross 
sections. According to Table~\ref{tbl:singlexsecsys}, we find that the 
cross sections at $\w<\gev{1.1}$ and $\w>\gev{2.0}$ are 
\SI{2.1}{\percent} and \SI{6.6}{\percent} of the cross section at 
$1.1<\w<\gev{2.0}$, respectively.

%=================================================== Table_II
\begin{table}[tbp]
    \centering
    \caption{Cross sections in ultraperipheral $\unpAu$ collisions at
    $\snn=\gev{200}$.}
    \begin{ruledtabular}
	\begin{tabular}{cccc}
	\multicolumn{3}{c}{$p\textrm{Au}\to{nX}$\,($\eta>6.9$ and $\xf>0.4$)} &
	\multicolumn{1}{c}{$\pAupipin$} \\
	\cline{1-3}     \cline{4-4}
	$<\gev{1.1}$ & $1.1$--$\gev{2.0}$ & $>\gev{2.0}$ & $1.25$--$\gev{2.0}$ \\
	$\mb{0.6}$   & $\mb{27.4}$        & $\mb{1.8}$   & $\mb{6.2}$          \\
	\end{tabular}
	\end{ruledtabular}
\label{tbl:singlexsecsys}
\end{table}

(2) The contribution of the two-pion production $\gppipin$ appears above 
the threshold energy $\w\approx\gev{1.25}$. The UPC cross section in 
Table~\ref{tbl:singlexsecsys} is calculated using the $\twomaid$ 
model~\cite{MAID2pi}, where the $\eta$ and $\xf$ limits are not applied 
to neutrons. Comparing UPCs leading to two-pion production, $\mb{6.2}$ 
present in Table~\ref{tbl:singlexsecsys}, with those leading to single 
pion production, $\mb{41.7}$ present in Table~\ref{tbl:singlexsec}, the 
former amounts to \SI{14}{\percent} to the latter cross section. 
According to the discussions in (1) and (2), we find that UPCs at 
$1.1<\w<\gev{2.0}$ leading to single neutron and pion production 
dominantly contribute to the single spin asymmetry for neutrons.

(3) Effects of nonzero $Q^2$ to single spin asymmetry in UPCs are tested 
by comparing the total cross sections and $d\sigma/d\Phi$ distributions 
between $Q^2=0$ and $Q^2 \neq 0$. For the nonzero $Q^2$ values, we use 
$Q^2=\SI{6e-4}{\giga\electronvolt^{2}}$ in $\pAu$ collisions and 
$Q^2=\SI{2e-3}{\giga\electronvolt^{2}}$ in $\pAl$ collisions. In both 
collisions, the cross section for forward neutron production at 
$Q^2\neq0$ is at most $\SI{2}{\percent}$ larger than those at $Q^2=0$. 
Because $d\sigma/d\Phi$ is proportional to $1+P_2\cos\Phi T(\theta_\pi)$ 
and $T(\theta_\pi)$ is a function of $Q^2$, the $d\sigma/d\Phi$ 
distribution is modified by $Q^2$ depending on $\cos\Phi$. Accordingly, 
$\an^{\textrm{UPC}(\unpAu)}$, obtained from $\langle T(\theta_\pi) 
\rangle$ averaged over $\w$ and $\theta_\pi$, at 
$Q^2=\SI{1e-3}{\giga\electronvolt^{2}}$ is $\sim\SI{10}{\percent}$ 
smaller than that at $Q^2=0$.

The model uncertainties discussed in this subsection are summarized in
Table~\ref{tbl:sys}.

%---------------------------------------------------- Fig_4

%=================================================== Table_III
\begin{table}[tbp]
    \caption{Summary of uncertainties in the UPC MC simulation.}
    \begin{ruledtabular}
    \begin{tabular}{l l l}
      (1) Energy range  & $\sigma_{<\gev{1.1}}/\sigma_{1.1-\gev{2.0}}$ & $\SI{2.1}{\percent}$ \\
                        & $\sigma_{>\gev{2.0}}/\sigma_{1.1-\gev{2.0}}$ & $\SI{6.6}{\percent}$ \\
      (2) Two-pion production & $\sigma_{\pAupipin}/\sigma_{\pAupin}$  & $\SI{14}{\percent}$  \\
      (3) $Q^2$ range   & $\sigma_{Q^2 \neq 0}/\sigma_{Q^2 = 0}$       & $<\SI{2}{\percent}$  \\
                        & $\an^{Q^2 \neq 0}/\an^{Q^2=0}$               & $-\SI{10}{\percent}$ \\
    \end{tabular}
    \end{ruledtabular}
    \label{tbl:sys}
\end{table}

\subsection{Simulation results in $\pAl$ collisions at $\sqs=\gev{200}$}
\label{sec:p+Al}

%--- cross section
Total cross sections for UPCs and hadronic interactions in $\pAl$ collisions are
summarized in Table~\ref{tbl:singlexsec}. The UPC cross section is
$\sigma_{\textrm{UPC}(\pAl)}=\mb{0.7}$ which is $\sim\SI{8}{\percent}$ of
$\sigma_{\textrm{HAD}(\pAl)}=\mb{8.3}$, where UPCs in $\pAl$ collisions are
highly suppressed compared with those in $\pAu$ collisions due to $\propto Z^2$.

%--- ds/dz
Figure~\ref{fig:dsdxf}(c) shows $d\sigma/d\xf$ for UPCs (dashed [red] line) and
hadronic interactions (solid [black] line).
We find UPCs leading to subdominant contribution to the $d\sigma/d\xf$
distribution at $\xf<0.95$.

%--- ds/dphi
Finally, Fig.~\ref{fig:dsdxf}(d) compares $d\sigma/d\Phi$ between UPCs (dashed
[red] line) and hadronic interactions (solid [black] line).
Although the UPC cross section is about $\SI{8}{\percent}$ of the
hadronic-interactions cross section, the large positive asymmetry of UPCs
eventually compensates the small negative asymmetry of hadronic interactions.

%
% ----- Discussions -----
%
\section{Discussions}\label{sec:discussions}

\begin{figure}[!htb]
  \includegraphics[width=1.0\linewidth]{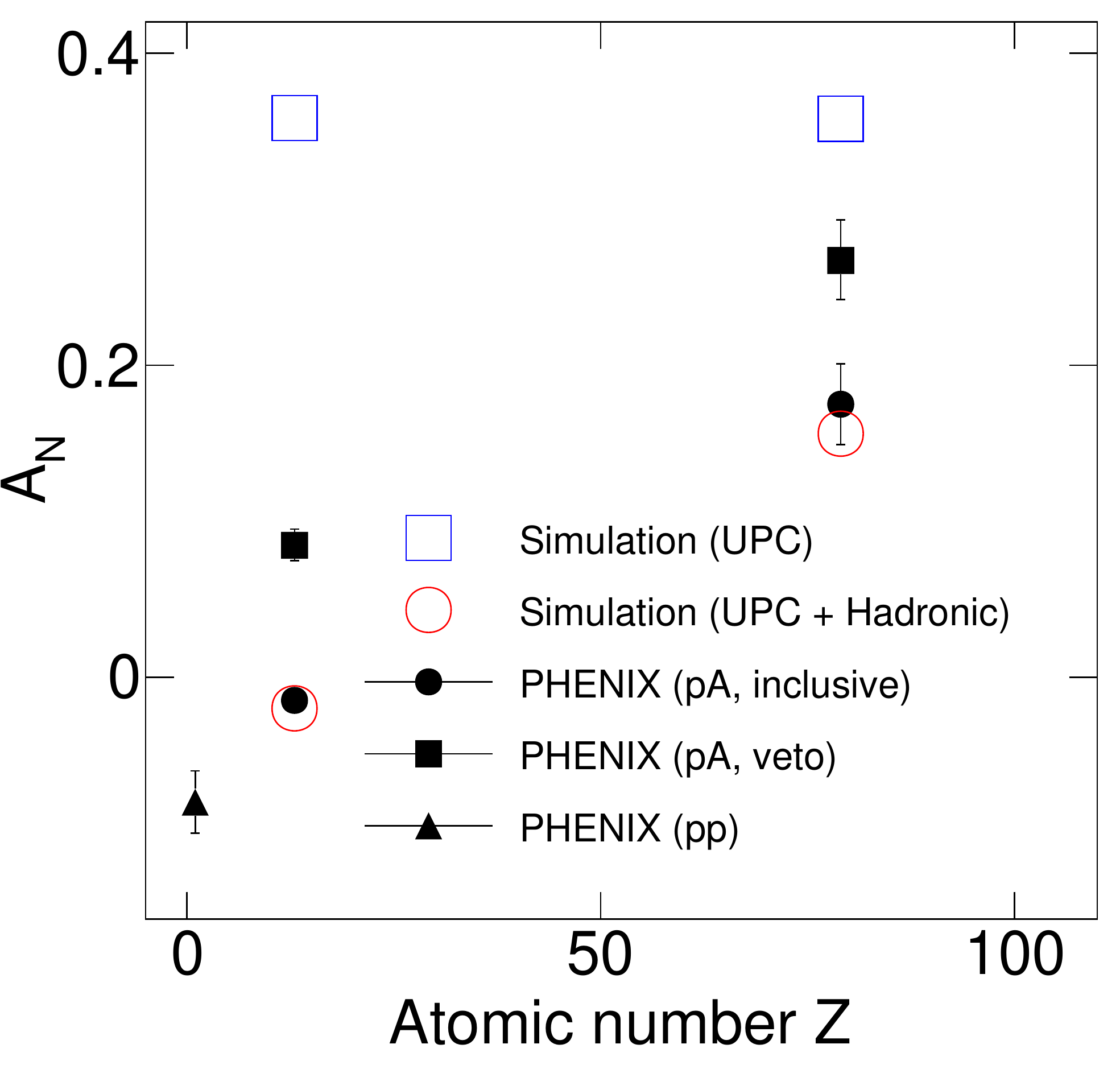}
  \caption{Transverse single spin asymmetry $\an$ of forward neutron. Filled
  black marker indicates the PHENIX results. Open [red] circle and open [blue]
  squares indicate the asymmetry obtained by the sum of UPCs and hadronic
  interactions and by only hadronic interactions, respectively.}
  \label{fig:AN}
\end{figure}

We compare the simulation results with the observed $\an$ values in $\pAl$ and
$\pAu$ collisions at $\snn=\gev{200}$.  Figure~\ref{fig:AN} shows $\an$ as a
function of the atomic number $Z$ in $\pp$ (for reference), $\pAl$ and $\pAu$
collisions.

Filled [black] circles indicate the $\an$ values inclusively measured by the
PHENIX zero-degree calorimeter~\cite{PHENIXPrelim}, where the neutron rapidity
and $\xf$ ranges are limited by $6.8<\eta<8.8$ and $\xf>0.4$, respectively.
These values can be compared with open [red] circles indicating the sum of UPCs
and hadronic interactions MC simulations, denoted $\an^{\textrm{UPC+HAD}}$.
These are written as
\begin{equation}
\an^\textrm{UPC+HAD}=
\frac{\sigma_\textrm{UPC}\an^{\textrm{UPC}} + \sigma_\textrm{HAD}\an^{\textrm{HAD}}}
{\sigma_\textrm{UPC} + \sigma_\textrm{HAD}},
\label{eq:antot}
\end{equation}
since
\begin{equation}
\frac{d\sigma_\textrm{UPC}}{d\Phi} + \frac{d\sigma_\textrm{HAD}}{d\Phi}
\propto 1+\cos\Phi \an^{\textrm{UPC+HAD}}.
\label{eq:dsdPhitot}
\end{equation}
For the MC simulation results (open [red] circles and open [blue] squares), the
neutron rapidity and $\xf$ region limits, $6.8<\eta<8.8$ and $\xf>0.4$, are also
taken into account to be consistent with the PHENIX measurements.
In $\pAl$ collisions, we obtain $\an^{\textrm{UPC+HAD}}=-0.02$ which is
consistent with the PHENIX result $\an=-0.015 \pm 0.005$. In $\pAu$ collisions,
we have $\an^{\textrm{UPC+HAD}}=0.16$ that can be understood by that UPCs,
having large positive $\an^{\textrm{UPC}}$ and a cross section
$\sigma_\textrm{UPC} \approx \sigma_\textrm{HAD}$, significantly contribute to
the inclusive $\an^{\textrm{UPC+HAD}}$ value that are evident in
Fig.~\ref{fig:dsdxf} and Table~\ref{tbl:singlexsec}.
Note that a model uncertainty in $\an^{\textrm{UPC+HAD}}$, estimated by taking
account of nonzero $Q^2$ discussed in Sec.~\ref{sec:$p$$+$Au_sys}, amounts
\SI{10}{\percent}.

Filled [black] squares in Fig.~\ref{fig:AN} are the $\an$ values measured 
by the PHENIX zero-degree calorimeter requiring a veto on the beam-beam counters 
(BBCs) covering $3.0<|\eta|<3.9$~\cite{RHICBBC}.  Because a nucleus in UPCs 
coherently scatters with a proton and thus does not generate underlying 
particles, such a BBC veto effectively selects UPC-rich events.
In $\pAu$ collisions, the PHENIX data with the BBC veto has larger $\an$ than
the inclusive PHENIX data (filled [black] circle).
This indicates that the fraction of UPCs in the PHENIX data is enhanced at a
certain level by the BBC veto, although the actual fraction is not presently
reported.
Conversely in $\pAl$ collisions, the PHENIX data with the BBC veto provides
$\an=0.085$ which is far smaller than $\an=0.27\pm0.03$ in $\pAu$ collisions.
A possible inference for the difference is that the fraction of hadronic
interaction is still sizable in the PHENIX data in $\pAl$ collisions even though
UPC-rich events are preferentially selected by the BBC veto.
If pure UPC data is experimentally available, the $\an$ values may approach
$\an^{\textrm{UPC}}$ in both $\pAl$ and $\pAu$ collisions.
Note that $\an^{\textrm{UPC}}$ is same between $\pAl$ and $\pAu$ collisions,
since $\an^{\textrm{UPC}}$ depends only on the $\gp$ interactions which are
common between $\pAl$ and $\pAu$ collisions.

%---------------------------------------------------- Fig_5

%
% ----- Conclusions -----
%
\section{Conclusions}\label{sec:conclusions}

It is demonstrated that ultraperipheral $\pA$ collisions have large $\an$ for
forward neutrons using the MC simulation framework developed for this study. 
The present UPC simulation comprised the following two parts; first, the
simulation of the virtual photon flux was performed by the \textsc{starlight}
event generator and, second, the simulation of the $\gppin$ interaction followed
the differential cross sections predicted by $\onemaid$ unitary isobar model. In
the $\gp$ interaction, the target asymmetry $T(\theta_\pi)$ was appropriately
treated.
According to the MC simulations, we found UPCs in $\pA$ collisions leading to
$\an^{\textrm{UPC}(\unpA)}=0.36$. Concerning forward neutron production of $\pA$
hadronic interaction, the simulation model used an one-pion exchange model and
the Glauber model. The single spin asymmetry was effectively taken in account by
multiplying $d\sigma_{\pAnX}/d\omegan$ with $1+\cos\Phi
\an^{\textrm{HAD}(\unpA)}$ where $\an^{\textrm{HAD}(\unpA)}=-0.05$. Combining
the differential cross sections of UPCs and hadronic interactions, we simulated
the $\xf$ and $\Phi$ distributions for inclusive forward neutrons. The $\an$
values for inclusive neutrons at $6.8<\eta<8.8$ and $\xf>0.4$ were predicted as
$-0.02$ and $0.16$ in $\pAl$ and $\pAu$ collisions, respectively. These were
consistent with the recently reported PHENIX results.
The PHENIX data with the BBC veto in $\pA$ collisions had larger $\an$ than the
inclusive PHENIX data, but were smaller than $\an^{\textrm{UPC}}$.
This indicated that requiring the BBC veto enhanced the fraction of UPCs in the
PHENIX data at a certain level, although the actual fraction was presently
unreported.

For future analyses, we plan to extend the present simulation framework to
include the contribution of the two-pion production $\gppipin$. This would
provide a more accurate description of $\an$ of forward neutrons.  Another
extension would be to investigate the possible interference between
electromagnetic and hadronic interactions, which is known as Coulomb-nuclear
interference.

% ----- Acknowledgements -----
\section*{Acknowledgments}\label{sec:acknowledgements}

The author appreciates fruitful discussions with Y.~Akiba, Y.~Goto and
I.~Nakagawa.

%
%---- Bibliography ----
%


\begin{thebibliography}{99}
\bibitem{PHENIXPrelim}
	I. Nakagawa \etal, (PHENIX Collaboration),
	The First Transverse Single Spin Measurement in High Energy Polarized Proton-Nucleus Collision at the PHENIX experiment at RHIC,
	J. of Phys. Conf. Ser. \textbf{736}, 012017 (2016).
\bibitem{PHENIXNeutron}
	A. Adare \etal, (PHENIX Collaboration),
	Inclusive cross section and single transverse spin asymmetry for very forward neutron production in polarized $p$$+$$p$ collisions at $\sqs=\gev{200}$,
	Phys. Rev. D \textbf{88}, 032006 (2013).
\bibitem{RHICZDC}
	C. Adler \etal,
	The RHIC zero degree calorimeters,
	Nucl. Instrum. Meth. \textbf{A 470} (2001) 488.
\bibitem{Kopeliovich2}
	B. Z. Kopeliovich, I. K. Potashnikova, and Iv\'an Schmidt,
	Single transverse spin asymmetry of forward neutrons,
	Phys. Rev. D \textbf{84}, 114012 (2011).
\bibitem{Kopeliovich3}
    B. Z. Kopeliovich, I. K. Potashnikova, and Iv\'an Schmidt,
	Leading Neutrons From Polarized Proton--Nucleus Collisions,
	arXiv:1611.07365.
\bibitem{Bertulani1}
	C. A. Bertulani and G. Baur,
	Electromagnetic processes in relativistic heavy ion collisions,
	Phys. Rep. \textbf{163} (1988) 299.
\bibitem{Bertulani2}
	C. A. Bertulani, S. R. Klein, and J. Nystrand,
	Physics of Ultra-Peripheral Nuclear Collisions,
	Ann. Rev. Nucl. Part. Sci. \textbf{55} (2005) 271.
\bibitem{Mitsuka}
	G. Mitsuka,
	Forward hadron production in ultra-peripheral proton–heavy-ion collisions at the LHC and RHIC,
	Eur. Phys. J. C \textbf{75}, 614 (2015).
\bibitem{STARLIGHT}
	S. R. Klein, J. Nystrand, J. Seger, Y. Gorbunov, and J. Butterworth,
	STARlight: A Monte Carlo simulation program for ultra-peripheral collisions of relativistic ions,
	Comput. Phys. Comm. \textbf{212} (2017) 258.
\bibitem{STARLIGHTcode}
	\textsc{starlight} webpage, \url{https://starlight.hepforge.org/}.
\bibitem{MAID07}
	D. Drechsel, S. S. Kamalov, and L. Tiator,
	Unitary isobar model --MAID2007,
	Eur. Phys. J. A \textbf{34}, 64.
\bibitem{Weizsacker}
	C. von Weizs\"{a}cker,
	Radiation emitted in collisions of very fast electrons,
	Z. Physik \textbf{88} (1934) 612.
\bibitem{Williams}
	E. J. Williams,
	Nature of the high-energy particles of penetrating radiation and status of ionization and radiation formulae,
	Phys. Rev. \textbf{45} (1934) 729.
\bibitem{Drechsel}
	D. Drechsel and L. Tiator,
	Threshold pion photoproduction on nucleons,
	J. Phys. G: Nucl. Phys. \textbf{18}, 449 (1992).
\bibitem{ISR}
	W. Flauger and F. M\"onnig,
	Measurement of inclusive zero-angle neutron spectra at the CERN ISR,
	Nucl. Phys. B \textbf{109}, (1976) 347.
\bibitem{ZEUS1}
	S. Chekanov \etal, (ZEUS Collaboration),
	Study of the pion trajectory in the photoproduction of leading neutrons at HERA,
	Phys. Lett. B \textbf{610} 199 (2005).
\bibitem{Gribov}
	V. N. Gribov,
	Interaction of gamma quanta and electrons with nuclei at high-energies,	
	Sov. Phys. JETP \textbf{30} (1970) 709.
\bibitem{Glauber}
	R. J. Glauber and G. Matthiae,
	High-energy scattering of protons by nuclei,
	Nucl. Phys. \textbf{B 21} (1970) 135.
\bibitem{Guzey}
    V. Guzey and M. Strikman,
    Proton–nucleus scattering and cross section fluctuations at RHIC and LHC,
	Phys. Lett. B \textbf{633} (2006) 245.
\bibitem{Kopeliovich1}
	B. Z. Kopeliovich, I. K. Potashnikova, Ivan Schmidt, and J. Soffer,
	Damping of forward neutrons in$p$$+$$p$collisions,
	Phys. Rev. D \textbf{78}, 014031 (2008).
\bibitem{H1}
	F.D. Aaron \etal, (H1 Collaboration),
	Measurement of leading neutron production in deep-inelastic scattering at HERA,
	Eur. Phys. J. C \textbf{68} (2010) 381.
\bibitem{PDG}
    K. A. Olive \etal~(Particle Data Group),
    Chin. Phys. \textbf{C 38} (2014) 090001.
\bibitem{ZEUS2}
	S. Chekanov \etal, (ZEUS Collaboration),
	Leading neutron production in $e^+p$ collisions at HERA,
	Nucl. Phys. B \textbf{637}, 3 (2002).
\bibitem{MAID2pi}
	A. Fix and H. Arenh\"ovel,
	Double pion photoproduction on nucleon and deuteron,
	Eur. Phys. J. A \textbf{25}, 115 (2005).
\bibitem{RHICBBC}
	M. Allen \etal,
	PHENIX inner detectors,
	Nucl. Instrum. Meth. \textbf{A 499} (2003) 549.
\end{thebibliography}
\end{document}